\documentclass[%
 aip,
rsi,
 amsmath,amssymb,
 reprint%
]{revtex4-1}

\usepackage{graphicx}
\usepackage{dcolumn}
\usepackage{bm}
\usepackage{float}
\usepackage{graphicx}
\usepackage{physics}
\usepackage{amsmath}
\usepackage[colorlinks=true,citecolor=blue,linkcolor=blue]{hyperref}
\usepackage{amssymb}
\usepackage{amsthm}
\usepackage{amsfonts}
\usepackage{braket}

\let\oldciteauthor=\citeauthor
\def\citeauthor#1{\hypersetup{citecolor=black}\oldciteauthor{#1}}
\let\oldcite=\cite
\def\cite#1{\hypersetup{citecolor=blue}\oldcite{#1}}

\begin{document}

\preprint{AIP/123-QED}

\title{Hybrid spin-superconducting quantum circuit mediated by deterministically prepared entangled photonic states}

\author{Kayleigh Mathieson}
\affiliation{Nano Scale Transport Physics Laboratory, School of Physics, University of the Witwatersrand, Private Bag 3, WITS 2050, Johannesburg, South Africa \\
}

\author{Somnath Bhattacharyya}
\email{somnath.bhattacharyya@wits.co.za}
\affiliation{Nano Scale Transport Physics Laboratory, School of Physics, University of the Witwatersrand, Private Bag 3, WITS 2050, Johannesburg, South Africa \\
}
\affiliation{National University of Science and Technology “MISiS”, Leninski Avenue 4, 119991 Moscow, Russia \\
}

\date{1 October 2019}

\begin{abstract}
\small{
In hybrid quantum systems a controllable coupling can be obtained by mediating the interactions with dynamically introduced photons. We propose a hybrid quantum architecture consisting of two nitrogen vacancy center ensembles coupled to a tunable flux qubit; that are contained on the transmission line of a multimode nonlinear superconducting coplanar waveguide resonator with an appended Josephson mixing device. We discuss using entangled propagating microwaves photons, which through our nonlinear wave-mixing procedure are made into macroscopically distinct quantum states. We use these states to steer the system and show that with further amplification we can create a similar photonic state, which has a more distinct reduction of its uncertainty. Furthermore, we show that all of this leads to a lengthened coherence time, a reasonable fidelity which decays to 0.94 and then later increases upward to stabilize at 0.6 as well as a strengthened entanglement.}
\end{abstract}

\maketitle

\small

Hybrid quantum systems have emerged as a potential solution, due to the properties which the interface between the different components can provide as an open quantum system.\cite{Xiang2013,Kurizki2015} Nitrogen vacancy center ensembles (NVEs) have been of importance due to their ability to be coupled and their stability in an open system.\cite{Song2015,Maleki:18,Maleki:19,PhysRevA.97.012312,PhysRevA.97.012329,PhysRevApplied.10.024011} Various studies have been done involving them collectively coupled to a flux qubit (FQ).\cite{Marcos2010,Qiu2014,PhysRevB.87.144516} This coupling can be extended beyond the strong coupling regime to ultrastrong domains.\cite{Stassi_2016} But, with the addition of more qubits the system would not be completely robust.

We use a modified superconducting coplanar waveguide resonator (CWR) as a multimode microwave photon quantum bus,\cite{Liu2016a,Tholén,Simoen,PhysRevApplied.10.044019} to which we apply quantum reservoir engineering, to create two-mode entangled microwave fields as the interaction medium. The microwave fields can be affected by the decoherence from the surrounding qubits, which make them less effective in transferring states within the system. One way to strengthen both the system and the microwave fields and make them less susceptible to dissipation is through transforming them into nonclassical states,\cite{PhysRevB.76.064305,PhysRevLett.121.123604,PhysRevA.91.013834} such as a coherent superposition state or Schr$\ddot{\text{o}}$dinger cat state.\cite{Arenz_2013} We look at the degree of squeezing applied and attempt to extend the work of \citeauthor{Minganti2016}\cite{Minganti2016} and \citeauthor{Hacker2019}\cite{Hacker2019}, which suggested the inherent squeezing of coherent cat states and a method of deterministically generating cat states.\cite{PhysRevA.71.063820} We look at a Schr$\ddot{\text{o}}$dinger cat-like state, which in our system. This state has been explored because of its ability to suppress noise, which can lead to more precise measurements, such as in quantum enhanced sensing.\cite{PhysRevA.93.033859} 

In our study, we look at cases involving a systematic creation and distribution of coherent and squeezed coherent macroscopically distinct states by the multimode parametric waveguide\cite{doi:10.1063/1.4933265,Zhu2017,Lo2015,PhysRevLett.120.093601,PhysRevLett.101.253602}, and compare their functions and the resulting system metrics to make a comparison between them. 

\begin{figure}
    \centering 
    \includegraphics[height=3cm, width=7cm]{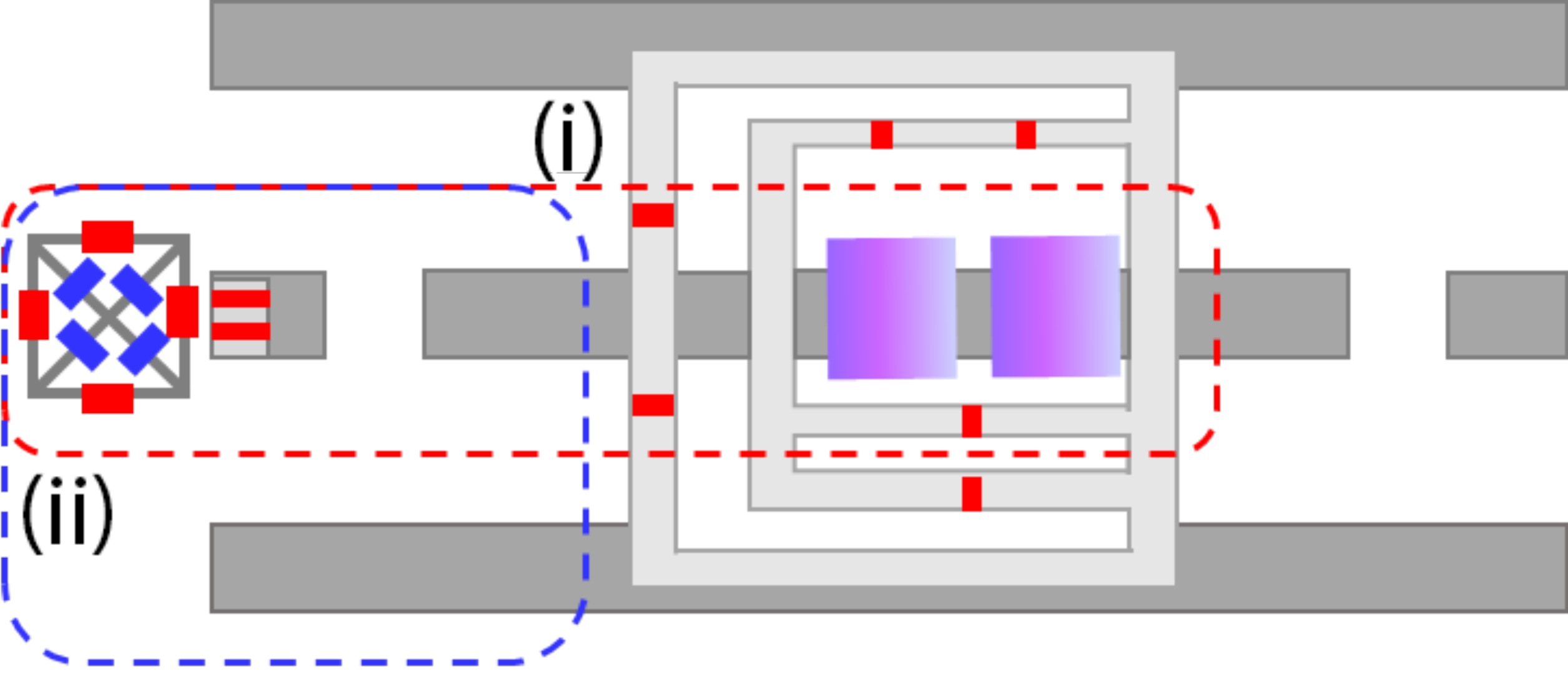}
    \caption{The quantum circuit. (i) The mechanism which creates the (squeezed) cat states. (ii) The part which transfers the entanglement to the NVEs.}
    \label{fig:circuit}
\end{figure}

\smallbreak

The system under consideration as illustrated in Figure 1, contains two nitrogen-vacancy center ensembles and a superconducting flux qubit coupled to a superconducting coplanar waveguide resonator with a Josephson mixing device. The non-degenerate four-wave mixing process taking place involves the conversion of the two pump modes into two signal modes. We represent this with the Hamiltonian $H_{w} = \hbar g_{w} (c_{1}^{+} c_{2} + c_{2}^{+} c_{1})(c_{3}^{+} c_{4} + c_{4}^{+} c_{3})$, where $c_{1}$ and $c_{2}$ are the pump operators, $c_{3}$ and $c_{4}$ are the signal operators and $g_{w}$ is a coupling constant. The superconducting microwave resonator with frequencies $\omega_{r_{1}}$ and $\omega_{r_{2}}$ are described as a harmonic oscillator and with the addition of the four-wave mixing device, which provides parametric amplification it becomes an effective parametric oscillator\cite{PhysRevLett.120.093601}. Its Hamiltonian is $H_{r}$ $=$ $\omega_{r_{1}} (a_{1}^{\dagger}a_{1}^{\dagger}$ $+$ $a_{1}a_{1})$ $+$ $\omega_{r_{2}} (a_{2}^{\dagger}a_{2}^{\dagger}$ $+$ $a_{2}a_{2})$, in which we have $a_1$, $a_2$ and $a^{\dagger}_1$, $a^{\dagger}_2$ as the respective annihilation and creation operators of the microwave fields of the resonator. 

The microwave photons also have a Rabi frequency $\Omega_{t}$. In our scheme the resonator has an added superconducting quantum interference device (SQUID) loop, to make its frequencies $\omega_{r_{1}}$, $\omega_{r_{2}}$, $\Omega_{1}$ and $\Omega_{2}$ tunable. The flux qubit has four Josephson junctions and an additional $\alpha$ loop with a DC SQUID to make it gap tunable. Its Hamiltonian is $H_{fq}$ $=$ $ \frac{\hbar}{2} (\delta_{z} \sigma_{z}$ $+$ $\delta_{x} \sigma_{x})$, where $\sigma_{z}$ and $\sigma_{x}$ denote the Pauli operators in the basis of the clockwise $\ket{{\circlearrowright}}$ and anticlockwise $\ket{{\circlearrowleft}}$ persistent current states. $\delta_{z}$ $=$ $\epsilon (\Phi_{ext})$ is the energy bias and $\delta_{x}$ $=$ $\Delta (\Phi_{0})$ is the qubit tunneling splitting basis of the flux qubit. 

\begin{figure}
    \centering
    \includegraphics[height=7.2cm, width=8cm]{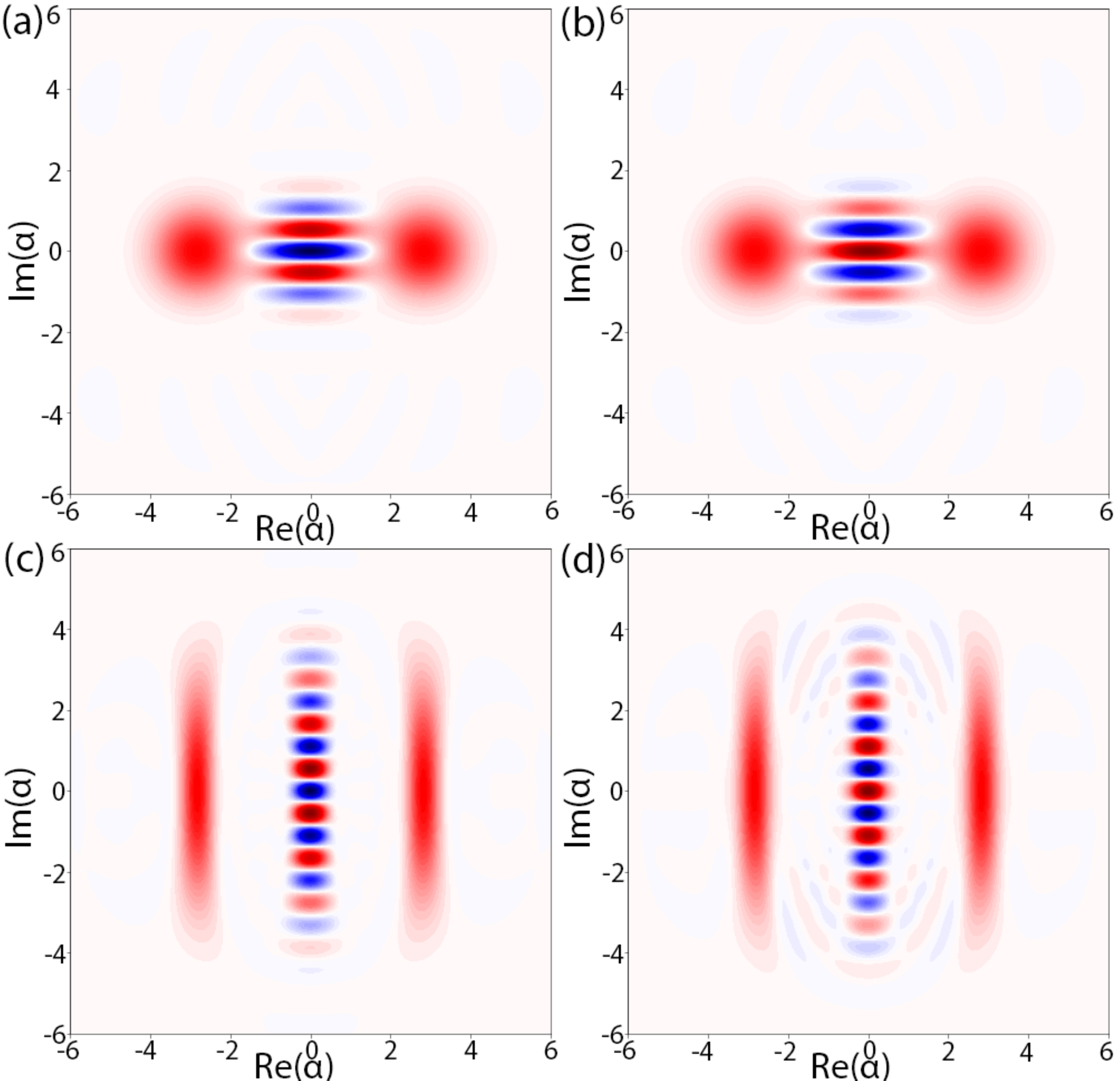}
    \caption{The Wigner functions with the odd cat states of amplitude $\alpha_{0}$ $=$ $2$ and squeezing $\phi$ $=$ $\pi$. Panels (a) and (b) show the function for the CSCS. Panels (c) to (d) show the function for the SSCS.}
    \label{fig:Wigner}
\end{figure}

\smallbreak

The nitrogen vacancy center ensemble Hamiltonian is $H_{nve}$ $=$ $\sum_{j=1}^{2} \sum_{i=1}^{N_{j}}[(D_{gs} (S_{z,i}^j)^{2}$ $+$ $g_{e} \mu_{B} B_{z} S_{z,i}^{j}]$. From the external magnetic field $B_{z}$ we get the zero-field splitting frequency $D_{gs}$ $\simeq$ $2.87$ GHz and the Zeeman splitting is $g_{e} \mu_{B} B_{z} S_{z,i}^{j}$.

$g_{e}$ is the ground-state Land\'{e} factor and $\mu_{B}$ is the Bohr magneton, there are $\mathcal{N}_{j}$ nitrogen vacancy centers in the $j$-th spin ensemble. We arrive at a reduced Hamiltonian in the bosonic basis, with frequency $\omega_j$ $=$ $D_{gs}$ $-$ $g_{e} \mu_{B} B_{z}$ and $\omega_j = 2.87$ GHz, which allows us to arrive at the full NVE Hamiltonian $H_{nve}$ $=$ $\sum_{j=1}^{2} \omega_{j} b_{j}^{+} b_{j}$. The full system Hamiltonian $H_{S}$ that describes the system is given by

\begin{align}
H_{S} &= \displaystyle\sum_{j=1}^{2} \omega_{j} b_{j}^{+} b_{j} + \delta_{x} \sigma_{x} + \omega_{r_{1}}  a_{1}^{\dagger} a_{1} + \omega_{r_{2}} a_{2}^{\dagger} a_{2} \nonumber \\
&+ \displaystyle\sum_{j=1}^{2} [G_{j}^{nvf} (b_{j}^{+} + b_{j}) \sigma_{z} + G_{j}^{nvr} (b_{j}^{+} a_{j} + a_{j}^{\dagger} b_{j})] \nonumber \\
&+ G^{fr}_{1} (a_{1} + a_{1}^{\dagger}) \sigma_{z} + G^{fr}_{2} (a_{2} + a_{2}^{\dagger}) \sigma_{z}
\end{align}

where $G^{nvf}$, $G^{fr}$ and $G^{nvr}$ are the coupling strengths between the NVE and the flux qubit, the flux qubit's persistent current and the resonator's photonic modes, and NVE and resonator's photonic modes respectively. $G^{nvf}$, $G^{nvr}$ are magnetic-dipole coupling strengths and $G^{fr}$ is the electric-dipole coupling strength. 

\smallbreak

We look at the system initially in the ground state, with the spin ensemble placed into a superposition state through a continuously applied $\pi/2$ pulse sequence. The flux qubit is prepared in its superposition state by the microwave lines of Rabi frequencies varying its magnetic flux. The modes $a_{1}$ and $a_{2}$ are shifted into an entangled coherent superposition state. The coherent state is taken in terms of the two-mode displacements from the vacuum state, by the operators $D (\alpha_1) = e^{\alpha_{1} a_{1}^{\dagger} - \alpha_{1}^{*} a_{1}}$ and $D (\alpha_2) = e^{\alpha_{2} a_{2}^{\dagger} - \alpha_{2}^{*} a_{2}}$. The coherent Schr$\ddot{\text{o}}$dinger cat state (CSCS) is of the form

\begin{equation}
\ket{CSCS(t){}^{\pm}} = \frac{1}{\sqrt{\mathcal{B}}} (\ket{\alpha_{1},\alpha_{2}} \pm e^{i \phi} \ket{-\alpha_{1},-\alpha_{2}}) 
\end{equation}

We look at the even and odd parity cat states with a normalization factor $\mathcal{B}$

\begin{equation}
\mathcal{B} = 2[1 + \exp(- 2\alpha_{1}^{2} \alpha_{2}^{2}) \cos \phi]
\end{equation}

We plot the Wigner quasi-probability distribution function \cite{PhysRev.40.749} to show the states of the microwave fields. They are obtained from the displacement parity operators $P_{1}$ with $W(\alpha_{1},\alpha_{2}) = \frac{2}{\pi} D(\alpha_{1}) P_{1} D(\alpha_{2})$. The CSCS is shown by both (a) as the even cat state and (b) with the odd cat state in Figure 2. The odd cat state displays a higher amplitude of destructive interference than the even cat state has of constructive interference in the positive direction.

\begin{figure}
    \includegraphics[height=6.2cm, width=8cm]{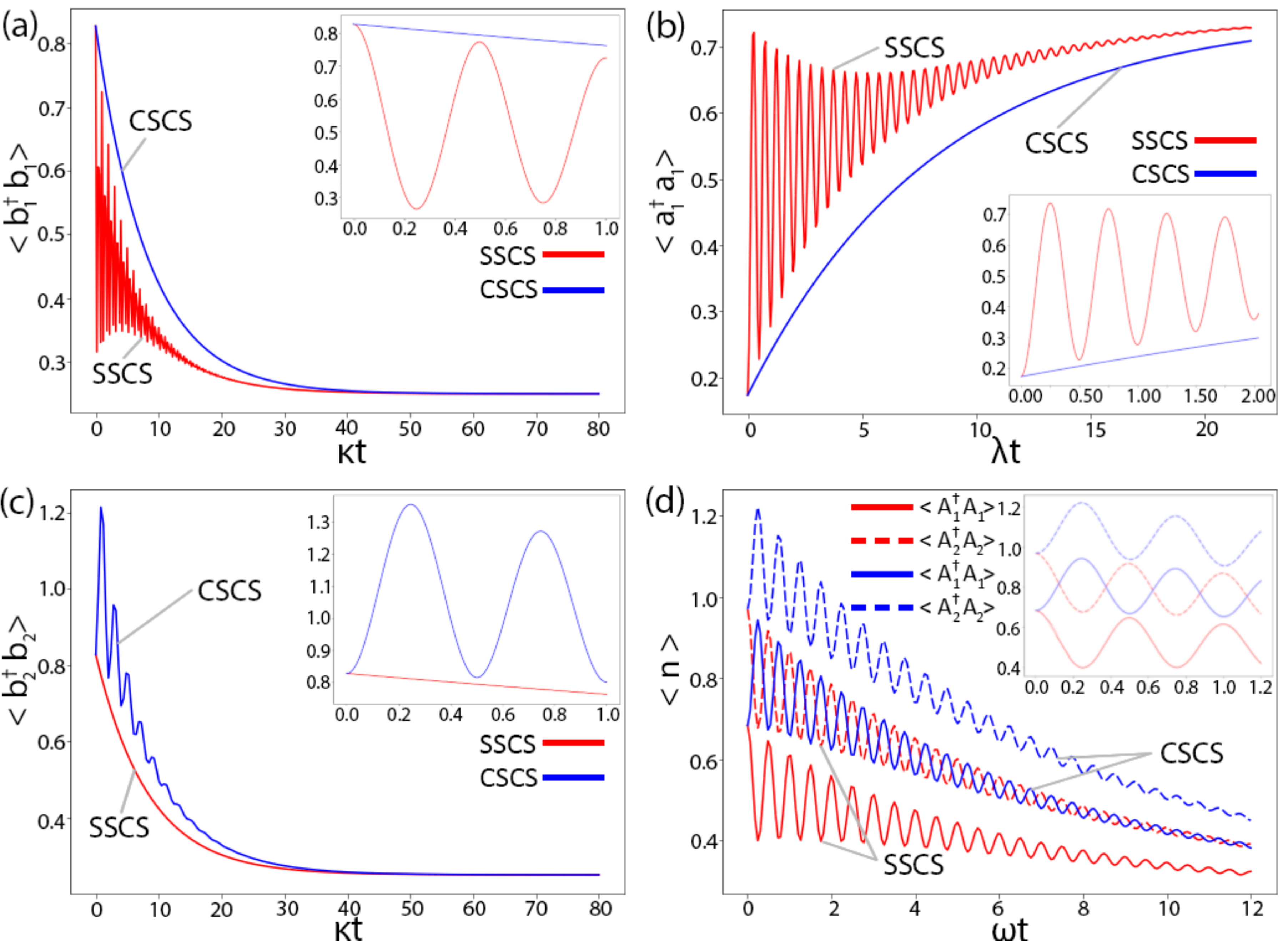}
    \caption{The expectation values for the population of states as a function of the dimensionless time parameter $\kappa t$, $\lambda t$ and $ut$. [(red) with the squeezed cat state as an initial state of the system and its evolution. (blue) with the coherent cat state as an initial state of the system and its evolution (or as denoted below in the following figures as RB)]. Panels (a) and (c) show the evolution of the excitation state of the first and second NVEs. The insets in (a) and (c) show the evolution under briefer time instances, being 1.0 and 2.0 respectively. Panels (b) shows the evolution of different states of the field modes. Then in Panel (d) the main graph shows the evolution of the population of states of the field modes in terms of the positive and negative superposition of states with them. The whole lines are for the positive superposition and similarly the dashed lines are for the negative superposition. The inset shows their evolution in a shorter time interval.}
    \label{fig:Plots2}
\end{figure}

\smallbreak

With the two-mode squeezed coherent state (SSCS), we take its squeezing operator as 

\begin{equation}
S (\ket{\xi}) = \exp (\xi a_{1}^{\dagger} a_{2}^{\dagger} - \xi^{*} a_{1} a_{2})
\end{equation}

where $\xi$ $=$ $r e^{i \phi}$. We define the operators $d_{1} = \mu a_{1} + \nu a_{1}^{\dagger}$ and $d_{2} = \mu a_{2} + \nu a_{2}^{\dagger}$, such that $\beta_{1} = \mu \alpha_{1} + \nu \alpha_{1}^{*}$ and $\beta_{2} = \mu \alpha_{2} + \nu \alpha_{2}^{*}$.  
They can then be turned into the squeezed Schr$\ddot{\text{o}}$dinger cat state (SSCS) as

\begin{align}
\ket{SSCS(t)^{\pm}} = \frac{1}{\sqrt{\hat{\mathcal{B}}}} (\ket{\beta_{1},\beta_{2}} \pm e^{i \phi} \ket{-\beta_{1},-\beta_{2}})
\end{align}

\smallbreak

where the normalization factor $\hat{\mathcal{B}}$ is given as

\begin{equation}
\hat{\mathcal{B}} = 2 \bigg[1 + \exp(\frac{\Xi -  4\alpha_{1}^{2} \alpha_{2}^{2} \cos^{2} \varphi}{(\mu - \nu)(\mu^{*} - \nu^{*})}) \cos \phi \bigg]
\end{equation}

where $\Xi = \Big([\alpha_{1}(\abs{\nu}^{2} + \abs{\mu}^{2} -2\mu\nu^{*}) - \alpha_{1}^{*}(\abs{\mu}^{2} - \abs{\nu}^{2} - 2\nu\mu^{*})] [\alpha_{2}(\abs{\nu}^{2} + \abs{\mu}^{2} -2\mu\nu^{*}) - \alpha_{2}^{*}(\abs{\mu}^{2} - \abs{\nu}^{2} - 2\nu\mu^{*})]\Big)$. We obtain the Wigner function for the squeezed Schr$\ddot{\text{o}}$dinger cat state, with the squeezed parity operator $P_{2}$ with $W(\beta_1,\beta_2) = \frac{2}{\pi} \ket{SC_{1}} P_{2} \ket{SC_{2}}$. The SSCS is shown by (c) and (d) in Figure 2. Their state is similar to that of the CSCS, with an additional phase difference. They have more numerous and dispersed points of interference as well as higher amplitudes in both parities for the even and odd states respectively. 

\smallbreak

The nitrogen vacancy center ensembles experience dephasing at the rate $\gamma_{j}$ and $\{j$=$1,2\}$ where we take $\gamma_1$ $=$ $\gamma_2$. $\gamma_{j}$ includes some of the residual effects of the ensemble's inhomogeneous broadening. The rate of dephasing in the flux qubit is given by $\Lambda$, the photonic modes in the coplanar waveguide resonator dephase at the rate $\kappa$. The detuning of the NVE is $\Delta_{nv}$ $=$ $\omega_{r_{j}}$ $-$ $\omega_{j}$ and of the flux qubit is $\Delta_{f}$ $=$ $\omega_{r_{j}}$ $-$ $\lambda$. We implement a unitary transformation $U$ $=$ $\exp[-i (\omega_{r_{1}} (a_{1}^{\dagger}a_{1}^{\dagger} + a_{1}a_{1})) (\omega_{r_{2}}(a_{2}^{\dagger}a_{2}^{\dagger} + a_{2}a_{2}))t]$, to take the system into the interaction rotating frame regime with a Hamiltonian

\begin{align}
H_{int} =& \displaystyle\sum_{j=1}^2 [G^{nvf}_j (b_j^{+} + b_j) \sigma_{z} + G^{nvr}_j (b_j^{+}a_{j} + a_{j}^{\dagger} b_j)] \nonumber \\
&{} + G^{fr}_{1} (a_{1}^{\dagger 2} e^{-i \omega_{q} t} + a_{1}^{2} e^{i \omega_{q} t}) \sigma_{z} \nonumber \\
&{} + G^{fr}_{2} (a_{2}^{\dagger 2} e^{-i \omega_{q} t} + a_{2}^{2} e^{i \omega_{q} t}) \sigma_{z}   
\end{align}

The Quantum Master Equation (QME) for the system in the coherent state representation is given as 

\begin{align}
\dot{\rho}(t) =&{} -i [H_{int},\rho] + \displaystyle\sum_{j=1}^{2} \mathcal{L}_{b_{j}}\rho + \mathcal{L}_{\sigma_{-}}\rho
+ \displaystyle\sum_{k=1}^{2} \mathcal{L}_{a_{k}}\rho
\end{align}

The whole expression for the above Liouvillian $\mathcal{L}$ is given in the Supplementary. We consider this in terms of just the lindblad dissipator $\mathcal{D}[o]\rho$ $=$ $2 o \rho o^{\dagger}$ $-$ $o^{\dagger} o \rho$ $-$ $\rho o^{\dagger} o$. This is extended with another QME

\begin{align}
\dot{\rho} (t) =&{} -i [H_{int}, \rho{t}] + \gamma_{1} \mathcal{D}[b_{1}]\rho \nonumber \\
&{}+ \gamma_{2}\mathcal{D}[b_{2}]\rho + \Lambda \mathcal{D}[\sigma_{\pm}]\rho \nonumber \\
&{}+ \kappa (\bar{n}_{th} + 1) \mathcal{D} [A_{1}] \rho + \kappa \bar{n}_{th} \mathcal{D} [A_{1}^{\dagger}] \rho \nonumber \\
&{}+ \kappa (\bar{n}_{th} + 1) \mathcal{D} [A_{2}] \rho + \kappa \bar{n}_{th} \mathcal{D} [A_{2}^{\dagger}] \rho   
\end{align}

We take the system under similar limits of experimental feasibility$\cite{PhysRevLett.100.077401,PhysRevLett.103.200404,PhysRevLett.110.250503,GaoJ,Lindstr_m_2007}$, and undertake a quantum simulation with values in the range of typical experimental parameters. $\bar{n}_{th}$ $=$ $({e^{\hbar \omega /k_{B} T} - 1})^{-1}$ corresponds to the thermal excitation number at the temperature $T$ and with frequency $\omega$. For numerical generality we look at a general case and as such we consider the frequencies to be the same among the different components. We take that $T$ $=$ $0.5$ Kelvin and limit the number $\mathcal{N}$ of NVC spins in the NVEs to $10^{2}$. We assign the coupling parameters as $G^{nvf}$ $=$ $2$ GHz, $G^{nvr}$ $=$ $0.05$ GHz and $G^{fr}$ $=$ $0.5$ GHz. The dissipative system parameters are $\gamma_{1}$ $=$ $0.08$, $\gamma_{2}$ $=$ $0.08$, $\Lambda$ $=$ $0.5$ and $\kappa$ $=$ $0.02$. 

\begin{figure}
    \centering
    \includegraphics[height=5cm, width=7cm]{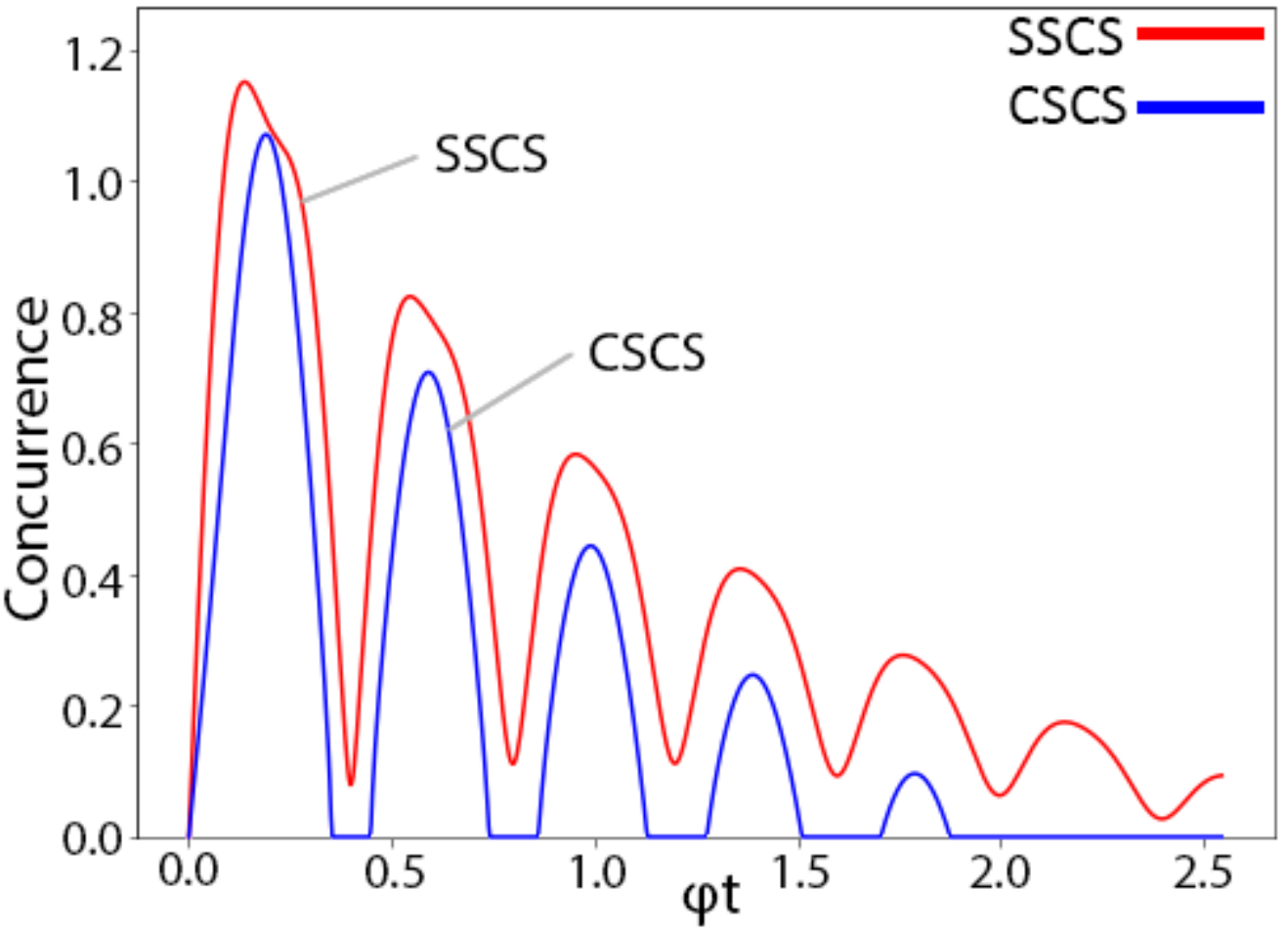}
    \caption{The evolution of the concurrence of the system with the dimensionless time parameter $\phi t$ $=$ $2.5$ [with RB].}
    \label{fig:Plots13}
\end{figure}

\begin{figure}
    \centering
    \includegraphics[height=5cm, width=7cm]{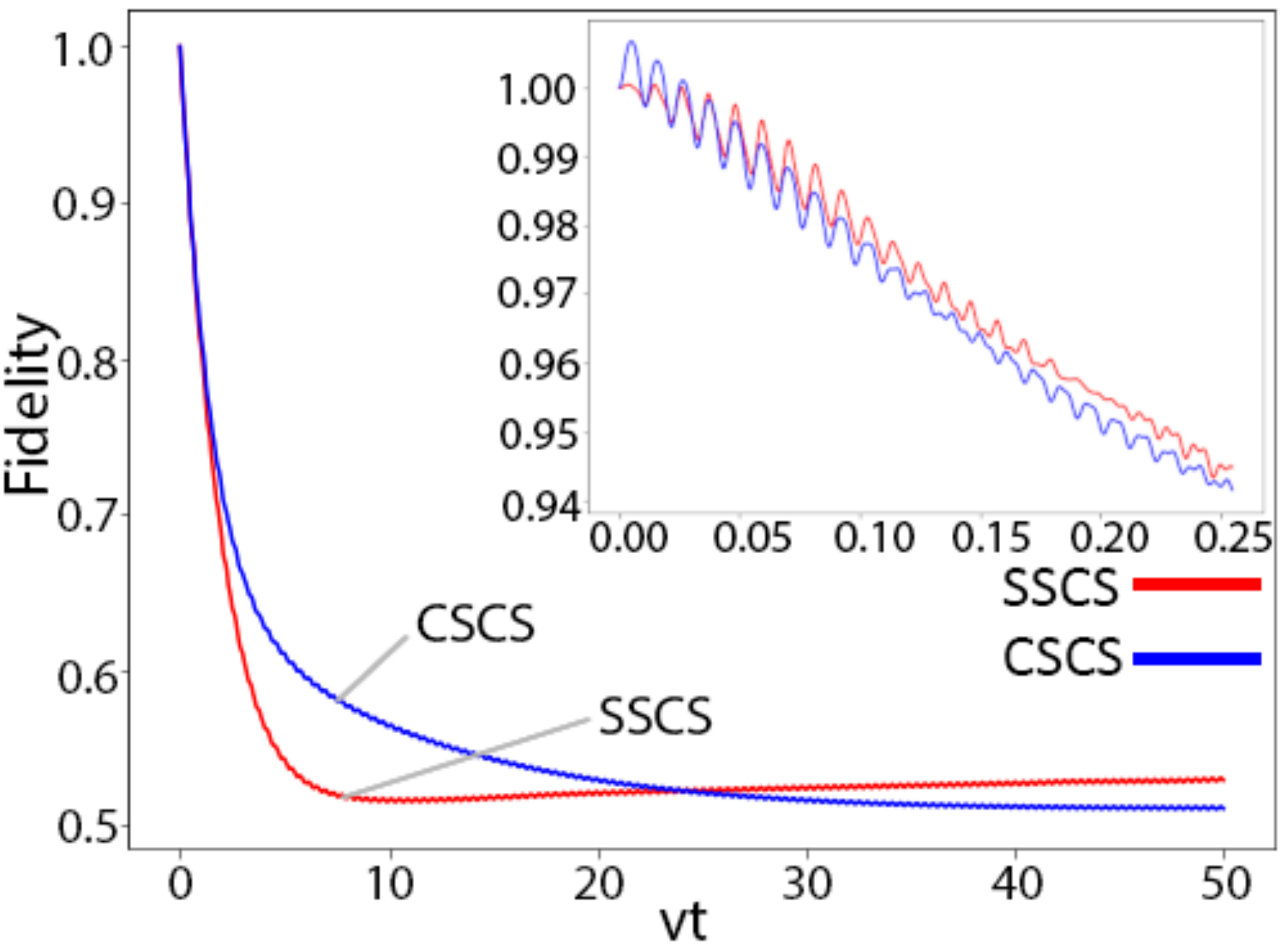}
    \caption{The evolution of the fidelity of the system with different dimensionless time parameters $wt$ and $vt$ [with RB]. In the main plot $vt$ $=$ $0.25$ and in the inset $wt$ $=$ $50$.}
    \label{fig:Plots14}
\end{figure}

\smallbreak

To show the evolution of the cat states and their effects in the system we have looked at the time evolution of the expectation values for its local state operators. In the figures the prominent decline or incline all lead to an eventual final steady probability. In Figure 3 we present the evolution of the expectation values for the NVEs as shown by (a) and (c) and also the expectation values for the field modes which are given as (b) and (d). As can be seen from the mentioned calculations, in (a) there is a steady decline in the probability, until it reaches a specific point at approximately $wt$ $=$ $30$ for the SSCS and $wt$ $=$ $40$ for the CSCS. Both the CSCS and SSCS lead to a probability is less than $0.3$. (b) shows the probability as increasing with time, as the field mode is in its excited state, although they both reach a steady point of approximately $0.7$. (c) reaches a similar conclusion to (a) although it is considering the second NVE, it has the same coherence time as (a). Plane (d) shows the time evolution of related expectation values representing the positive and negative superposition of the field modes states. From this we can gather that the negative superposition with the CSCS yields a higher initial probability than the other states and similarly with the negative superposition SSCS there is also a higher likelihood than with its positive counterpart. Although, they all transition to the same final probability of approximately $0.3$. The inset shows this behaviour in a shorter time scale, with the negative superposition states yielding a higher probability. 

\smallbreak

As defined by Wooters concurrence \cite{PhysRevLett.70.1895,An_Min_2003} in which the concurrence is taken as $C (\rho)$ $=$ $\max(0,$ $\lambda_{1}$ $-$ $\lambda_{2}$ $-$ $\lambda_{3}$ $-$ $\lambda_{4})$. Where the $\lambda_{i}$'s (with $\lambda_{1}$ being the maximum) are the eigenvalues from the following matrix $\sqrt{\sqrt{\rho} \tilde{\rho} \sqrt{\rho}}$ with $\tilde{\rho}$ $=$ $(\sigma_{y} \otimes \sigma_{y}) \rho^{*} (\sigma_{y} \otimes \sigma_{y})$. In Figure 4 we evaluate the concurrence as a function of time for the system. The figure and its fluctuating nature indicates that the cat states entangle them in an oscillating manner. It shows that initially the system is affected and its entanglement is degraded by the decoherence, however in time $>$ $2.5$ the cat states lead to the progression to a steady state of concurrence $0.7$. In the case presented at first the SSCS has a concurrence higher than that of the CSCS, this can be understood by its additional interference, which leads to stronger and therefore more resilient entanglement in the system. After the initial decrease the concurrence rises first with the SSCS and then with the CSCS to their final steady state. The resulting stability  is due to the result of the propagation of the cat state in the system. 

\smallbreak

To examine how the system and qubit dissipation affect each of the photonic states we have looked at the fidelity of the interacting system. The fidelity is defined by $F = \big[ \Tr{\sqrt{\sqrt{\rho} \sigma \sqrt{\rho}}} \big]^{2}$, where the ideal case with no dissipation ($\sigma$) is compared to our case which takes into account dissipation ($\rho$). As shown in Figure 5, the fidelity of the two systems initially follow the same trend, decreasing in value until a minimum fidelity is reached at approximately 0.5. Interestingly, although the SSCS fidelity decreases more rapidly, after reaching the minimum value, it shows an upturn and eventually increases to values greater than those in the CSCS case. Although the decrease in fidelity is quite substantial ($\sim$ 0.5), this value seemingly saturates indicative of the systems reaching a steady state. However the increase in the fidelity of the SSCS case is not expected, it may be possible when considering the model as a system of collective spins, essentially spin centers of different strength and stabilities. Upon excitation, the spin alignment of the individual elements interact and eventually reach a steady state (saturating fidelity). The crossing point can be interpreted as the crossover from a metastable state (fluctuating spins) to a steady state (coherent spin state). The phase difference also is contributing to the more rapid drop off the SSCS relative to that of the CSCS and the eventual occurrence of their cross-over when $vt = 24$. In fact it is well known that such spin interactions are the leading causes of decoherence that limit the performance of quantum information systems based on NV centers from \citeauthor{PhysRevA.97.012312} \cite{PhysRevA.97.012312}. 

In summary we have derived a model for a hybrid quantum system utilizing the coplanar waveguide as a multimode resonator, which generate and send photonic states through the system. Through our numerical simulations, we show that we can generate entanglement in this system, through the NVEs taking on the states of the photons which they interact with, through the state transfer. We look at Schr$\ddot{\text{o}}$dinger cat states and squeezed Schr$\ddot{\text{o}}$dinger cat states as the mediators of the interactions in the system. The photonic cat states dissipate as a result of their interaction with the qubits. The presented results suggest that the additionally squeezed SSCS leads to a stronger entanglement and higher fidelity. Our results suggest at a way of creating more robust and controllable quantum systems with states that are feasible to construct experimentally and which contribute to this effort, thereby overcoming the constraints which hinder their realization and limit their ability in implementation. We recommend further study regarding the utility of these states in similar systems.

\smallbreak

See the supplementary material for additional information regarding the derivations and an extension of the work which looks at the system components, specifically the NVEs and the flux qubit separately and their respective evolution of their concurrence and fidelity. 

\smallbreak

SB is very thankful to C. Coleman, S. Mukhin, A. Karpov and Y. Hardy for valuable discussion. SB thanks CSIR-NLC and acknowledges financial support from the Ministry of Education and Science of the Russian Federation in the framework of the Increased Competitiveness Program of NUST-MISiS (grant No. К3-2018-043).

\bibliography{main.bib}

\end{document}